\newcommand{\hn}[2]{\hat{n}^{#1}_{#2}}                   
\newcommand{\ha}{\hat{a}^{}}                             
\newcommand{\hadag}{\hat{a}^\dagger}                     
\newcommand{\hc}{\hat{c}^{}}                             
\newcommand{\hcdag}{\hat{c}^\dagger}                     
\newcommand{\sigz}{\hat{\sigma}_{z}}                            %
\newcommand{\hpsi}{\hat{\Psi}}                           
\newcommand{\hpsidag}{\hat{\Psi}^{\dagger}}                
\newcommand{\hphi}{\hat{\Phi}}                           
\newcommand{\hphidag}{\hat{\Phi}^\dagger}                
\newcommand{\bx}{{\bf x}}                                
\newcommand{\bk}{{\bf k}}                                
\newcommand{\bp}{{\bf p}}                                
\newcommand{\ket}[1]{|{#1}\rangle}                       
\newcommand{\bra}[1]{\langle {#1}|}                      

\documentclass[12pt]{iopart}

\usepackage{graphicx}

\begin{document}

\title{Collective decoherence of cold atoms coupled to a Bose-Einstein condensate}
\author{M.A. Cirone$^1$, G. De Chiara$^2$, G. M. Palma$^3$ and A. Recati$^4$}
\address{$^1$Dipartimento di Scienze Fisiche ed Astronomiche, Universit\`{a} degli Studi di Palermo,\\
via Archirafi 36, I-90123 Palermo, Italy}
\address{$^2$ Grup d'{\`O}ptica, Departament de F{\'i}sica, Universitat Aut{\`o}noma
de Barcelona,\\ E-08193 Bellaterra, Spain}
\address{$^3$ NEST - CNR - INFM and Dipartimento di Scienze Fisiche ed Astronomiche, \\ Universit\`{a} degli Studi di Palermo,
via Archirafi 36, I-90123 Palermo, Italy}
 \address{$^4$  Dipartimento di Fisica, Universit\`a di Trento
, CNR-INFM BEC Center, I-38050 Povo, Trento, Italy, and Physik-Department, Technische Universit\"at M\"unchen,  D-85748
Garching, Germany}

\ead{massimo.palma@fisica.unipa.it}
\date{\today}

\begin{abstract}
We examine the time evolution of cold atoms (impurities) interacting with 
an environment consisting of a degenerate bosonic quantum gas. 
The impurity atoms differ from the environment atoms, being of a different species.   This  allows one to superimpose  two independent trapping potentials, each being effective only on one atomic kind, while transparent to the other.
When the environment is homogeneous and the impurities are confined in 
a potential consisting of a set of double wells, the system
can be described in terms of an effective spin-boson model, where the occupation of the
left or right well of each site represents the two (pseudo)-spin
states. 
The irreversible dynamics of such system is  here studied exactly, i.e., not in terms
of a Markovian master equation.
The dynamics of one and two impurities  
is remarkably different in respect of the standard decoherence of the spin - boson system. In particular we show:
({\em i}) the appearance of coherence oscillations, ({\em ii}) the presence of super and sub
decoherent states which differ from the standard ones of the spin boson model, and ({\em iii})
the persistence of coherence in the system at long times. We show that this behaviour is due
to the fact that the pseudospins have an internal spatial structure. We argue
that collective decoherence also prompts information about the correlation length
of the environment.
In a one dimensional configuration one can change even stronger the qualitative
behaviour of the dephasing just by tuning the interaction of the bath.

\end{abstract}
\maketitle


\section{Introduction}
The reasons of the great interest for  the physics of ultracold atoms in recent years
are manifold. On the one hand experimentalists have
reached an unprecedented control over the many-body atomic state with very stable optical potentials and by the use of Feshbach resonances which allow one to change the scattering length of the atoms \cite{Blochreview}. In this context the tremendous experimental results that have been achieved include: the observation of the superfluid-Mott insulator transition for bosons \cite{bloch-qpt}, one dimensional strongly interacting bosons in the Tonks-Girardeau regime \cite{tonks} and Anderson localization \cite{aspect,inguscio}.
On the other hand new experimental challenges come from different theoretical proposals for using this system for quantum information processing \cite{qip} and as a quantum simulator of condensed matter models (see for example \cite{toolbox, Sanpera, RMP} and references therein).

Not only can ultracold atoms simulate Hamiltonian systems, but such systems also offer a way to engineer non classical environments. Thanks to the flexibility of quantum gases, a broad range of regimes
of irreversible dynamics of open quantum systems and in particular of spin-boson systems can be explored
\cite{alessio, lehur1, lehur2, griessner, fleisch}. 

In the present paper we propose a new way in which an instance of the  spin - boson model \cite{ZwergerRMP}
can be realized with a suitable arrangement of interacting cold atoms. 
In particular we  analyse a system consisting of cold impurity atoms interacting
with a degenerate quantum gas of a different atomic species. This setup makes possible the
superposition of two independent trapping potentials,  each being effective on one
atomic species only, while transparent to the other. When the quantum gas is homogeneous
and the impurities are confined in a potential composed of double wells,
the system can be described in terms of an effective spin-boson model, where
occupations of the left or right well represent the two (pseudo)-spin states.
At variance with other setups, where the role of the pseudospin is played by the presence
or absence of one particle in a trapping well \cite{garcia}, by the
vibrational modes of a single well \cite{eric} or by internal electronic levels \cite{fleisch},
in our case each pseudospin has a spatial dimension,
namely the  separation between the two minima of the impurity double well.
This introduces an effective suppression of the decoherence due to low frequency
modes of the environment and leads to unusual and interesting phenomena, like
oscillations of coherence at finite times and the survival of coherence at long times.
Further novel features appear when one considers the irreversible collective decoherence of a systems of two impurities. 
In this case we still predict the existence of subdecoherent and of superdecoherent state,
but with the interesting fact that their role is exactly the opposite from what one
observes in conventional spin-boson systems. 
Further interesting features appear when one considers how the collective decoherence
rates change as a function of the impurities' separation and the effects of dimensionality
of the system.

In discussing our investigations, for the sake of simplicity
we shall consider an experimental setup where the impurity
atoms are trapped by a periodic (optical) lattice. We like to stress, however, that our
findings do not depend on the lattice properties (e.g., periodicity) but for the numerical
results. Other setups, such as microtraps on atom chips or quantum dots, just to mention
a few, can be equally envisaged.

\section{The Hamiltonian}

\begin{figure}
  \centering
\includegraphics[height=7cm]{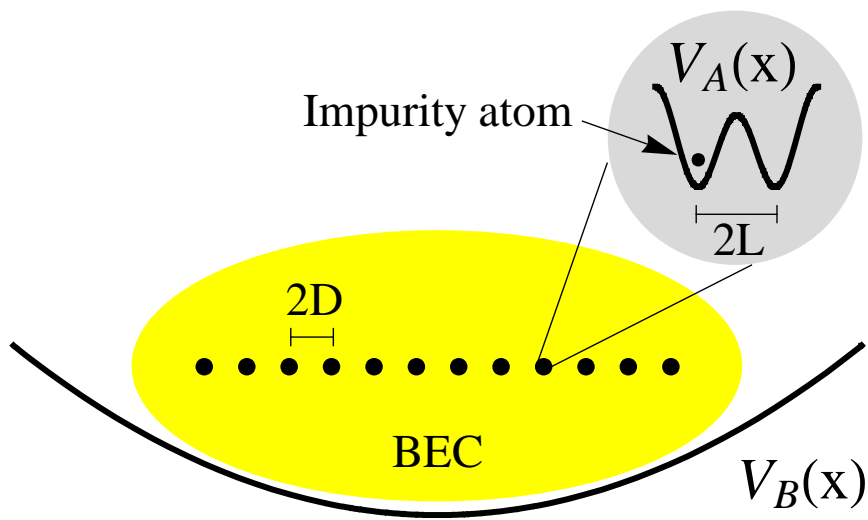}
  \caption{
A Bose-Einstein condensate (yellow region) confined in a shallow harmonic trap $V_B(x)$ interacts with cold impurity atoms each of which is trapped in a double well potential $V_A(x)$ (gray circle). The distance between two wells in the same trap is 2L and the distance between adjacent traps is 2D.}

  \label{fig1setup}
\end{figure}
Our system is composed of a cold quantum gas of bosonic atoms and a sample of
cold atoms separated from each other and immersed in the quantum gas.
In presenting our investigations, we shall use the words `reservoir', `bath' and
`environment' as synonyms to indicate the quantum gas, since its properties are
not the focus of the present paper.

The second-quantized form of the Hamiltonian of the impurities+bath system takes the form (see also Ref.\cite{jaksch2})

\begin{equation}
\hat{H}=\hat{H}_{\rm A}+\hat{H}_{\rm B}+\hat{H}_{\rm AB}
\label{hamil1}
\end{equation}
where

\begin{eqnarray}
\hat{H}_{\rm A} & = & \int \!\!\! d^3x \; \hpsidag(\bx) \left[
\frac{\bp_A^2}{2m_A}+V_A(\bx) \right] \hpsi(\bx)
\end{eqnarray}
is the Hamiltonian of atomic impurities, described by the field operator $\hpsi(\bx)$
in the trapping potential $V_A(\bx)$ which creates a set of double wells of size $2L$ and separated by a distance $2D$, see Fig.~\ref{fig1setup},

\begin{eqnarray} \hat{H}_{\rm B} & = & \int \!\!\! d^3x \;
\hphidag(\bx) \left[ \frac{\bp^2}{2m_B}+V_B(\bx)
+\frac{g_B}{2} \; \hphidag(\bx)\hphi(\bx) \right] \hphi(\bx)
\end{eqnarray}
is the Hamiltonian of the bath, composed of $N\gg 1$ bosons,
represented by the field operator $\hphi(\bx)$ and confined by a trapping potential
$V_B(\bx)$ and $g_B=4\pi \hbar^2
a_B/m_B$ is the boson-boson coupling constant, with $a_B$ the
scattering length of the condensate atoms, and

\begin{eqnarray}
\hat{H}_{\rm AB} & = &  g_{AB} \int \!\!\! d^3x \;
\hpsidag(\bx)\hphidag(\bx) \hphi(\bx)\hpsi(\bx)
\end{eqnarray}
describes the interactions between the impurities and the bath;
here $g_{AB}=2 \pi \hbar^2a_{AB}/m_{AB}$ is the coupling constant
of impurities-gas interaction, with $a_{AB}$ the scattering length
of the impurities-gas collisions and $m_{AB}=m_A m_B / (m_A+m_B)$
their reduced mass. 
Both impurity and bath atoms are described
in the second-quantized formalism. The field operator of the atomic
impurities

\begin{eqnarray}
\hpsi(\bx) = \sum_{i,p}  \ha_{i,p}\varphi_{i,p}^{}(\bx)
\label{wavepsi}
\end{eqnarray}
can be decomposed in terms of the real eigenstates
$\varphi_{i,p}(\bx)^{}$ of impurity atoms localized on the double well $i$ of
the potential $V_A(x)$ in the $p^{th}$ state,  with energy $\hbar
\omega_{i,p}$ and the corresponding  annihilation operator $\ha_{i,p}$ . 
We assume that the wavefunctions of different double wells have a negligible common support, i.e.,
$\varphi_{i,p}^{}(\bx)\varphi^{}_{j\neq i,m}(\bx) \simeq 0$ at any position $\bx$.

We treat the  gas of bosons following Bogoliubov's approach (see, for instance, Ref.\cite{trento})
and assuming a very shallow trapping potential $V_B(\bx)$, such that the bosonic gas
can be considered homogeneous. In the degenerate regime the bosonic field can be decomposed as

\begin{eqnarray}
\hphi(\bx) & = & \sqrt{N_0}\; \Phi_0(\bx) + \delta \hat{\Phi}(\bx) =
\sqrt{N_0}\; \Phi_0(\bx) + \sum_{\bk} \left( u_\bk(\bx)\hc_\bk - v^*_\bk(\bx) \hcdag_\bk \right)
\label{wavephi}
\end{eqnarray}
where $\Phi_0(\bx)$ is the condensate wave function (or order
parameter), $N_0<N$ is the number of atoms in the condensate and
$\hc_{\bk}$, $\hcdag_{\bk}$ are the annihilation and creation
operators of the Bogoliubov modes with momentum $\bk$.
For a homogeneous condensate $\Phi_0(\bx)=1/\sqrt{V}$, $V$ being the volume. Its Bogoliubov
modes

\begin{eqnarray}
u_\bk = \sqrt{\frac{1}{2}\left(\frac{\epsilon_\bk +
n_0g_B}{E_\bk}+1 \right)} \;
\frac{e^{i \bk \cdot \bx}}{\sqrt{V}}, \\
v_\bk = \sqrt{\frac{1}{2}\left(\frac{\epsilon_\bk +
n_0g_B}{E_\bk}-1 \right)} \; \frac{e^{i \bk \cdot \bx}}{\sqrt{V}}
\end{eqnarray}
have energy

\begin{eqnarray}
E_{\bk} & = & \left[2\epsilon_\bk n_0g_B +\epsilon_\bk^2
\right]^{1/2},
\label{Ek}
\end{eqnarray}
where $\epsilon_\bk=\hbar^2 k^2/(2m_B)$ and $n_0=N_0/V$ is the
condensate density. As one can see from (\ref{Ek}), low-energy excitations have
phonon-like (wave-like) spectrum, whereas high-energy excitations
have particle-like spectrum. The condition for wave-like
excitations is $\epsilon_\bk \ll n_0g_B$, i.e., $k \ll
4\sqrt{\pi n_0 a_B}$, or equivalently $k \ll 2m_B c_s/\hbar $, where
$c_s=\sqrt{n_0g_B/m}$ is the speed of sound at zero temperature.
Note that $\left| u_\bk \right|=1/\sqrt{V}$ and $\left| v_\bk
\right|=0$ describe the limiting case of $N\gg 1$ non-interacting
bosons, each with energy $E_\bk=\epsilon_\bk$.

Inserting Eqs.(\ref{wavepsi}) and (\ref{wavephi}) into the
Hamiltonian (\ref{hamil1}) we get

\begin{eqnarray}
\hat{H}_{\rm A} & = & \sum_{i,p} \hbar \omega_{i,p} \hadag_{i,p} \ha_{i,p}
\end{eqnarray}
for the impurities, 

\begin{eqnarray}
\hat{H}_{\rm B} & = & H_{\rm Cond}+\hat{H}_{\rm Bog}
\end{eqnarray}
for the quantum gas, with

\begin{eqnarray}
H_{\rm Cond} & = & N_0 \int \!\!\! d^3x \;
\Phi_0^*(\bx) \left[
\frac{\bp^2}{2m_B}+V^B(\bx)+\frac{g_B}{2}
\; N_0 |\Phi_0(\bx)|^2 \right] \Phi_0(\bx)
\label{hbath}
\end{eqnarray}
for the condensate and

\begin{eqnarray}
\hat{H}_{\rm Bog} & = & \sum_\bk E_\bk^{} \hcdag_\bk \hc_\bk
\end{eqnarray}
for the collective excitations (Bogoliubov modes)  of energy $E_{\bk}^{}$ in the condensate,
and

\begin{eqnarray}
\hat{H}_{AB} & = & g_{AB} \sum_i \sum_{p,q} \hadag_{i,p}
\ha_{i,q}
\left[ N_0 \int d^3x \varphi_{i,p}(\bx)\varphi_{i,q}(\bx) |\Phi_0(\bx)|^2+ \right. \nonumber \\
& & + \sqrt{N_0} \sum_{\bk} \hc_\bk \int d^3x \varphi_{i,p}(\bx)\varphi_{i,q}(\bx)
\left( \Phi_0^*(\bx) u_\bk(\bx) - \Phi_0(\bx) v_\bk(\bx) \right) \nonumber \\
& & +\left. \sqrt{N_0} \sum_{\bk} \hcdag_\bk \int d^3x \varphi_{i,p}(\bx)\varphi_{i,q}(\bx)
\left( \Phi_0(\bx) u_\bk^*(\bx) - \Phi_0^*(\bx) v_\bk^*(\bx) \right) \right]
\label{hint}
\end{eqnarray}
for the interaction Hamiltonian; the terms which are quadratic in the
Bogoliubov excitation operators $\hc, \hcdag$ give negligible contributions and
have been omitted.
The first term in (\ref{hint}) describes transitions between impurities'
vibrational states due to the condensate, whereas the
remaining terms describe similar transitions induced by the
collective excitations in the condensate.
In a homogeneous condensate, transitions between different vibrational
eigenstates of the impurities induced by the condensate are suppressed, while all vibrational states
$\varphi_{i,p}(\bx)$ get an energy shift $\delta \omega_{i,p}$,

\begin{equation}
g^{AB} N_0 \int \!\!\! d^3x \; |\Phi_0|^2(\bx) \varphi_{i,p}(\bx)\varphi_{i,q}(\bx) =
\cases{0 & for $p \neq q$ \\ n_0g^{AB}\equiv \delta \omega_{i,p} & for $p=q$ }
\end{equation}
so the contribution of the first term in (\ref{hint})
can be included in the definition of $\omega_{i,p}$.

In the limit of deep, symmetric wells in each double well and
separated by a high energy barrier, the tunneling between adjacent wells
is suppressed. In this regime the ground states $\varphi_{i,L}$ and
$\varphi_{i,R}$ of, respectively, the left and right
well of double well $i$ are well separated in space with vanishing spatial overlap,
their coupling to the excited states becomes negligible and the total Hamiltonian
further simplifies into

\begin{eqnarray}
\hat{H} & = & \sum_{i,} \sum_{p=L,R}
 \hbar \omega_{i,p} \hn{i}{p}+\sum_\bk
E_\bk^{} \hcdag_\bk \hc_\bk + 
 \sum_{i}\hbar \sum_{p=L,R} \sum_{\bk}
\left[ \Omega_{p,\bk}^i \hc_\bk + \Omega_{p,\bk}^{i*} \hcdag_\bk
\right] \hn{i}{p} \label{htot}
\end{eqnarray}
where we have defined the coupling frequencies

\begin{eqnarray}
\Omega_{p,\bk}^i \equiv \frac{g_{AB}\sqrt{n_0}}{\hbar}\left(
|u_\bk|-|v_\bk| \right) \int d^3x \; |\varphi_{i,p}(\bx)|^2 e^{i
\bk \cdot \bx}
\label{ome}
\end{eqnarray}
and $\hn{i}{p}\equiv \hadag_{i,p} \ha_{i,p}$ is the number
operator of impurities in the double well $i$ in the well $p=L,R$.

We consider the case where each double well is occupied by at most one impurity atom.
This allows us to describe the occupation of the left and right well of each site in terms of
pseudospin states. Introducing the Pauli operators as $\hn{i}{L}=(1-\sigz^i)/2$,
$\hn{i}{R}=(1+\sigz^i)/2$, the Hamiltonian (\ref{htot}) takes the form of the
independent boson model \cite{mahan}

\begin{eqnarray}
\hat{H} & = & \sum_\bk
E_\bk^{} \hcdag_\bk \hc_\bk + \frac{\hbar}{2} \sum_{\bk}
\left\{ \left[ \sum_{i} \left(\Omega_{R,\bk}^i-\Omega_{L,\bk}^i\right) \sigz^i +
\sum_{i} \left(\Omega_{R,\bk}^i+\Omega_{L,\bk}^i\right)\right]\hc_\bk \right. \nonumber \\
& & \left.  + \left[ \sum_{i}
\left(\Omega_{R,\bk}^{i*}-\Omega_{L,\bk}^{i*}\right) \sigz^i +
\sum_{i}
\left(\Omega_{R,\bk}^{i*}+\Omega_{L,\bk}^{i*}\right)\right]\hcdag_\bk
\right\} \label{hpseudo}
\end{eqnarray}
where a constant energy shift has been omitted. 
We note that spin-boson systems with larger spin values can be realized
in the same way with higher occupation of the double wells.

The effects due to quantum noise on coherent superpositions of states of a double well spin-boson 
hamiltonian have been analyzed in the markovian regime. In \cite{Micheli, Zurek, Savage} the effects
 of a cold atom reservoir has been analyzed, while \cite{Huang} has considered the effects of 
 scattered photons, taking into account also the role of the inter-well separation. 
 As we will show in the following section, for our system it is possible to carry a full analysis of the impurity dynamics, 
 going beyond the Markov approximation.


\section{Exact reduced impurities dynamics}

The dynamics due to the spin-boson Hamiltonian (\ref{hpseudo})
is amenable of an exact analytical solution and is characterized by decoherence
without dissipation \cite{massimo,quiroga,braun}.
The time-evolution operator $\hat{U}(t) = \exp \left[ -i\hat{H}t/\hbar \right]$
corresponding to the Hamiltonian (\ref{hpseudo}) can be factorized into
a product of simpler exponential operators,

\begin{eqnarray}
\hat{U}(t) & = & \exp \left[ -\frac{i}{\hbar}\sum_\bk E_\bk \hcdag_\bk
\hc_\bk t \right] \nonumber \\
& & \times \exp \left[ \sum_\bk \left( \sum_i A_\bk^i(t)\sigz^i+
\alpha_\bk^{}(t) \right) \hcdag_\bk  - \sum_\bk \left( \sum_i A_\bk^{i*}(t)\sigz^i+
\alpha_\bk^{*}(t) \right) \hc_\bk \right] \nonumber \\
& & \times \exp \left[ i\hbar^2\sum_{\bk}f_\bk (t) \Re\sum_{ij}
\frac{\left(\Omega_{R,\bk}^{i}-\Omega_{L,\bk}^{i}\right)
\left(\Omega_{R,\bk}^{j*}-\Omega_{L,\bk}^{j*}\right)}{4E_\bk^2}
 \sigz^i\sigz^j \right]
 \nonumber \\
& & \times \exp \left[ i\hbar^2 \sum_{\bk}f_\bk (t) \Re\sum_{i}
\frac{\left(\Omega_{R,\bk}^{i}-\Omega_{L,\bk}^{i}\right)
\sum_j \left(\Omega_{R,\bk}^{j*}+\Omega_{L,\bk}^{j*}\right)}{2E_\bk^2}
 \sigz^i \right]
 \nonumber \\
& & \times \exp \left[ i\hbar^2\sum_{\bk}f_\bk (t)
\frac{\sum_i\left(\Omega_{R,\bk}^{i}+\Omega_{L,\bk}^{i}\right)
\sum_j
\left(\Omega_{R,\bk}^{j*}+\Omega_{L,\bk}^{j*}\right)}{4E_\bk^2}
\right]
\label{equt}
\end{eqnarray}
where the functions

\begin{eqnarray}
f_\bk (t) & = & \frac{E_\bk}{\hbar}t-\sin \frac{E_\bk}{\hbar}t, \\
A_\bk^i(t) & = & \frac{\hbar \left( 1-e^{iE_\bk t/\hbar} \right)
}{2E_\bk}\left(\Omega_{R,\bk}^{i*}-\Omega_{L,\bk}^{i*}
\right), \\
\alpha_\bk^{}(t) & = & \frac{\hbar \left( 1- e^{iE_\bk t/\hbar}\right)}{2E_\bk}\sum_i
\left(\Omega_{R,\bk}^{i*}+\Omega_{L,\bk}^{i*}
\right), \;\;\;\;\;
\end{eqnarray}
have been introduced for ease of notation. Details of the derivation
of (\ref{equt}) for the time evolution operator are given in Appendix A.
As in this paper we are interested in the irreversible collective decoherence of the impurities we will focus our attention on the 
conditional displacement operator

\begin{eqnarray}
\hat{U}_D(t) & = & \prod_{\bk} \hat{U}_{\bk,D}(t), \\
\hat{U}_{\bk,D}(t) & \equiv & \exp \left[ \left( \sum_i A_\bk^i(t)\sigz^i+
\alpha_\bk^{}(t) \right) \hcdag_\bk  - \left( \sum_i
A_\bk^{i*}(t)\sigz^i+ \alpha_\bk^{*}(t) \right) \hc_\bk \right]
\end{eqnarray}
Indeed this operator is the one responsible of the decoherence of impurities as it induces entanglement between them and the reservoir.
Labeling the state of the impurities as
$\ket{\{n_p\}}=\ket{\{n_{1},n_{2},n_{3},\ldots\}}$
with $n_p=0,1$ denoting the presence of the atom, respectively, in the
left or right well, the matrix elements of reduced density operator of the impurities 
are

\begin{eqnarray}
\rho_{\{n_p\},\{m_p\}}(t) & = & \exp \left[ -\Gamma_{\{n_i\},\{m_i\}}(t)\right] \rho_{\{n_p\},\{m_p\}}(0)
\nonumber \\
& & \times \exp \left\{i
\Theta_{\{n_p\},\{m_p\}}(t) \right\}
\exp \left\{i \Xi_{\{n_p\},\{m_p\}}(t) \right\}
\exp \left\{i \Delta_{\{n_p\},\{m_p\}}(t) \right\}
\label{eqrho}
\end{eqnarray}
Assuming that each mode of the bosonic environment
is in a mixed state $\rho_\bk$ at equilibrium at temperature $T$ the decay exponent  contains all the information concerning the time dependence of the decoherence  process and takes the form
\begin{equation}
\Gamma_{\{n_i\},\{m_i\}}(t) = \hbar^2 \sum_\bk \frac{\left( 1-\cos \frac{E_\bk}{\hbar}t
\right)}{E_\bk^2} \left| \sum_i \left[m_i-n_i\right]\left(
\Omega_{R,\bk}^{i}-\Omega_{L,\bk}^{i} \right) \right|^2 \coth
\frac{\beta E_\bk}{2}
\label{gammas}
\end{equation}
with $\beta = 1/K_B T$. The phase factors $\Theta_{\{n_p\},\{m_p\}}(t)$, $\Xi_{\{n_p\},\{m_p\}}(t)$
and $\Delta_{\{n_p\},\{m_p\}}(t)$, whose specific form is given in appendix B, do not play any role in the decoherence \cite{Fazio}.
They contain however interesting information on the effective coupling between the pseudospins induced by the consensate and will be analysed in a future paper \cite{CDPR2} . 

\section{Results for the decoherence}

As mentioned in the Introduction, we shall assume that the impurity
atoms are trapped by an optical (super)lattice, whose form can be controlled
and varied in time with great accuracy \cite{anderlini,sebby}.
The coupling frequencies $\Omega_{p,\bk}^i$ are accordingly evaluated in Appendix C assuming
an optical lattice, with identical, double wells
in each site, and deep trapping of impurity atoms in their wells, with
identical confinement in each direction.
Atomic wave functions can then be approximated by harmonic oscillator
ground states of variance parameter $\sigma=\sqrt{\hbar/(m\omega)}$
\cite{mott}, where $\omega$ is the corresponding harmonic frequency.
As will be clear shortly, $\sigma$ acts as a natural cutoff parameter,
quenching the coupling with high frequency modes.

Specifically, we consider $^{23}$Na impurity atoms trapped in a far-detuned optical
lattice and a $^{87}$Rb condensate.
The condensate density is $n_0=10^{20}\, \rm{m}^{-3}$,
the lattice wavelength is $\lambda=600 \;{\rm nm}$,
and we have taken $2L=\lambda/2$ and $D=2L$. The depth of the optical lattice is
described by the parameter $\alpha \equiv V_0/E_R$, $V_0$ being the optical lattice
potential maximum intensity and $E_R=\hbar^2k^2/(2m)$ the recoil energy of impurity
atoms in the lattice; in our evaluations we put $\alpha=20$. Finally, we assume
$a_{AB}=55 a_0$ \cite{narb}, where $a_0$ is the Bohr radius, for the
scattering length of impurities-condensate mixtures. This parameter can be modified
in laboratory with the help of Feshbach resonances.

\subsection{Single  impurity decoherence}

We first examine the decoherence exponent of a single impurity

\begin{eqnarray}
\Gamma_0(t) \equiv \Gamma_{\{0\},\{1\}}(t) \equiv \hbar^2 \sum_\bk
\frac{\left( 1-\cos \frac{E_\bk}{\hbar}t \right)}{E_\bk^2}\coth \frac{\beta E_\bk}{2}
 \left| \Omega_{R,\bk}^{1}-\Omega_{L,\bk}^{1} \right|^2
\end{eqnarray}
Such quantity, that will be a useful benchmark in our analysis of the collective decoherence  of impurity pairs, shows already interesting features.
Assuming, from now on, that the condensate is at temperature $T=0$, we obtain
\begin{eqnarray}
\Gamma_0(t) =8 g_{AB}^2 n_0\sum_\bk \left( |u_\bk|-|v_\bk|\right)^2
e^{-k^2\sigma^2/2} \frac{\sin^2 \frac{E_\bk}{2\hbar}t
}{E_\bk^2}\sin^2 \left( \bk \cdot {\bf L} \right)
\end{eqnarray}
We note the dependence of $\Gamma_0 (t)$ on the length  ${\bf L}$, where  $2{\bf L}$ is the
distance between two wells within each site. The presence of the factor $\sin^2 \left( \bk \cdot {\bf L} \right) $ 
supresses the effect of the reservoir modes at small $\bk $. This is clearly understandable: environment modes whose 
wavelength is longer than ${\bf L}$  cannot ``resolve'' the spatially separated wells within each site. The consequences of this fact will be clear shortly. Replacing the sum over discrete modes to a continuum with the usual rule  $V^{-1}\sum_\bk \rightarrow (2\pi)^{-3} \int d\bk$,
choosing $x$ as azimuthal axis and using well known relations for Bogoliubov modes
\cite{castin}, we finally obtain
\begin{equation}
\Gamma_0^c(t) = \frac{2g_{AB}^2 n_0}{\pi^2}\int_0^\infty \!\!\!
dk \left[\; k^2 e^{-k^2\sigma^2/2}\frac{\sin^2
\frac{E_\bk}{2\hbar}t}{E_\bk\left( \epsilon_\bk + 2 g_B n_0 \right)}
 \right]  \left( 1-\frac{\sin 2kL}{2kL} \right) 
 \label{gammaz}
\end{equation}
The superscript $c$ is to remind us that we are dealing with impurities interacting with a condensate. For the special case of a bath of noninteracting bosons 
$ \Gamma_0^{n.i.}(t)$ is obtained from (\ref{gammaz}) simply imposing $g_B=0$ and $E_\bk=\epsilon_\bk$.
Let us point  out that the spectral density, which reads
\begin{equation}
J(\omega)\equiv
\sum_\bk |\Omega_{R,\bk}-\Omega_{L,\bk}|^2 \delta(\hbar\omega-E_\bk),
\label{specden}
\end{equation}
has a non trivial form, which at small frequencies, scales as $\omega^{d+2}$ for the interacting case, where $d$ is the dimensionality of the condensate, and as $\omega^{d/2}$ for the non interacting case. It is worth noticing that while the former case is always superohmic,
the latter one is subohmic, ohmic and superohmic depending on the dimensionality of the environment. Note that the high power in $J(\omega)$ is due to the fact that the bath has to ``resolve'' the structure of the impurity, formally again the factor $\sin^2 \left( \bk \cdot {\bf L} \right)$.
Furthermore, as already pointed out, no ``ad hoc'' cutoff frequency $\omega_c$ needs to be inserted 
but appears naturally in the decaying exponential of variance $\sigma$ in (\ref{gammaz}).

\begin{figure}
  \centering
  \includegraphics[height=6cm]{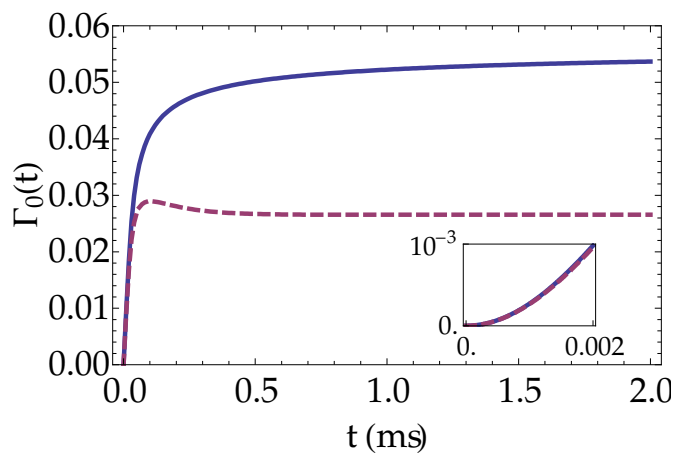}
  \caption{$\Gamma_0(t)$ vs. time for a single impurity atom interacting with free bosons
  (solid line) and with a bosonic condensate (dashed line) in three dimensions. The inset shows 
  $\Gamma_0(t)$ for very short times $0 \le t \le 2 \, \mu \rm{s}$.}
  \label{figgammazerocomparison}
\end{figure}
Fig.\ref{figgammazerocomparison} shows clearly that the
impurity maintains much of its coherence at long times. 
Such survival is due to the above mentioned suppressed effect of soft modes, which
are responsible of the long time behavior of $\Gamma_0 (t)$, and is more pronounced when the environment consists of 
a condensate than in the case of a reservoir consisting of free bosons. This can be intuitively described in terms of greater "stiffness" of the condensate 
whose Bogoliubov modes are less displaced by the coupled impurity. The condensate is
even able to give some coherence back to the impurity, since $\Gamma_0^c(t)$ is
not monotonic in time. Oscillations of coherence in spin-boson systems were predicted
in \cite{quiroga} (and even earlier, in a different context, in \cite{paz}).

We can distinguish three stages in the dynamics of the $\Gamma_0$'s. In the first
stage $\Gamma_0(t) \propto t^2$, as can be easily seen from a series expansion
of (\ref{gammaz}). This very short stage, shown in the inset of Fig.\ref{figgammazerocomparison}, corresponds to coherent dynamics. 
The second stage corresponds to a Markovian
behavior, i.e., $\Gamma_0(t) \propto t$, and lasts a few tens of microseconds.
Finally, in the third stage $\Gamma_0(t)$ saturates to a stationary value. 
This behaviour calls for particular caution
in treating an environment of (free or interacting) bosons as a Markovian reservoir
for atomic impurities immersed in it, which is clearly not the case in the present situation.

\begin{figure}
  \centering
  \includegraphics[width=16cm]{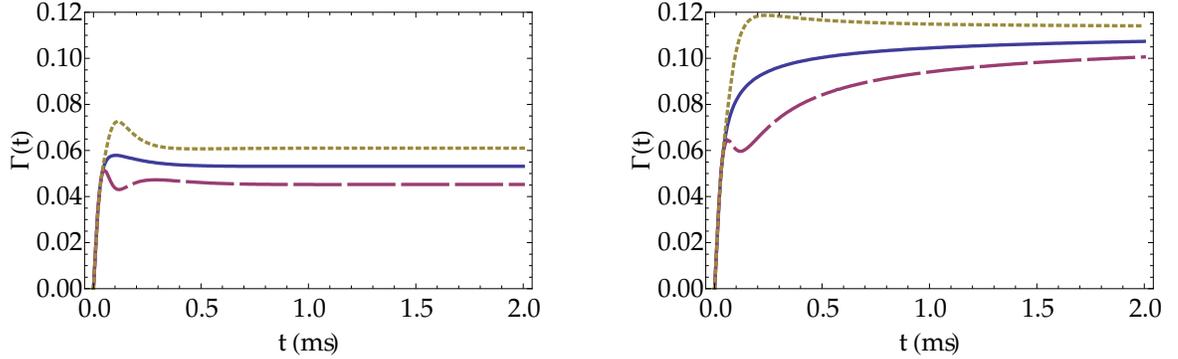}
  \caption{$\Gamma_1(t)$ (dashed line), $\Gamma_2(t)$ (dotted line), and $2\Gamma_0(t)$
  (solid line) vs. time for a pair of impurity atoms at a distance $2 D= 4 L$ (see text), immersed in
  a condensate (left) and in an environment of free bosons (right) in three dimensions.}
  \label{figgammaboth}
\end{figure}

\subsection{Collective decoherence of two impurities}

Decoherence of quantum systems in a common environment is characterized by collective decoherence.
It is well known that two  spins interacting with the same bosonic  reservoir with a spin - boson interaction Hamiltonian like the one discussed in this paper 
show sub -- and super -- decoherence \cite{massimo}. In simple words the decoherence rate of the two spins is not simply $2\Gamma_0(t)$ but, according to the initial state of the 
spins, much smaller or larger. In this final section of the present manuscript we analyse the specific features of collective decoherence in our system. 

For two pseudospins, three decoherence parameters appear in the density matrix elements
independently of the exact form of the impurities' state.
One is $\Gamma_0(t)$ and appears in elements such as $ \rho_{0,0;0,1}(t)$,
$\rho_{0,1;1,1}(t)$, etc., which corresponds to individual dephasing of each impurity atom; two more parameters $\Gamma_1(t)$ and $\Gamma_2(t)$
appear in elements such as
$\left| \rho_{0,0;1,1}(t) \right|=\exp \left[-\Gamma_1(t)
\right] \left| \rho_{0,0;1,1}(0) \right|$ and $\left| \rho_{0,1;1,0}(t)\right| =\exp
\left[-\Gamma_2(t) \right]\left| \rho_{0,1;1,0}(0)\right|$, and corresponds to decay of the coherences between states with the particles in the same or in the opposite side, respectively,
of the double well. For two pseudospins at distance $2\bf D=4L$, these two parameters are
\begin{eqnarray}
\Gamma_1(t) & \equiv & \Gamma_{\{0,0\},\{1,1\}}(t) =  \hbar^2 \sum_\bk \frac{\left( 1-\cos \frac{E_\bk}{\hbar}t
\right)}{E_\bk^2}\coth \frac{\beta E_\bk}{2}  \left| \left(
\Omega_{R,\bk}^{1}-\Omega_{L,\bk}^{1}+\Omega_{R,\bk}^{2}-\Omega_{L,\bk}^{2} \right) \right|^2 \nonumber\\
& = & 32 g_{AB}^2 n_0\sum_\bk \left(
|u_\bk|-|v_\bk|\right)^2 e^{-k^2\sigma^2/2} \frac{\sin^2
\frac{E_\bk}{2\hbar}t }{E_\bk^2}\sin^2 \left( \bk \cdot {\bf L} \right)
\cos^2 \left( \bk \cdot {\bf D} \right) 
\end{eqnarray}
\begin{eqnarray}
\Gamma_2(t) & \equiv & \Gamma_{\{1,0\},\{0,1\}}(t) =  \hbar^2 \sum_\bk \frac{\left( 1-\cos \frac{E_\bk}{\hbar}t
\right)}{E_\bk^2}\coth \frac{\beta E_\bk}{2}  \left| \left(
\Omega_{R,\bk}^{1}-\Omega_{L,\bk}^{1}-\Omega_{R,\bk}^{2}+\Omega_{L,\bk}^{2} \right) \right|^2 \nonumber \\
& = & 32 g_{AB}^2 n_0\sum_\bk \left(
|u_\bk|-|v_\bk|\right)^2 e^{-k^2\sigma^2/2} \frac{\sin^2
\frac{E_\bk}{2\hbar}t}{E_\bk^2}\sin^2 \left(\bk \cdot {\bf L} \right)
\sin^2 \left( \bk \cdot {\bf D} \right)
\end{eqnarray}
Calculations similar to those performed for $\Gamma_0$ give for a condensate environment

\begin{eqnarray}
\Gamma_1^c(t) & = &  \frac{2g_{AB}^2 n_0}{\pi^2}\int_0^\infty \!\!\!
dk \; k^2 e^{-k^2\sigma^2/2} \, \frac{\sin^2
\frac{E_\bk}{2\hbar}t }{E_\bk\left( \epsilon_\bk + 2 g_B n_0 \right)} \nonumber \\
&\times& \left(2 - 2\frac{\sin 2kL}{2kL}+ 2\frac{\sin2kD}{2kD}-\frac{\sin 2k(L+D)}{2k(L+D)}-\frac{\sin
2k(D-L)}{2k(D-L)} \right)  \nonumber\\
&\equiv&  2\Gamma_0(t) - \delta^c(t) \label{gammapl}\\
\Gamma_2^c(t) & = & \frac{2g_{AB}^2 n_0}{\pi^2}\int_0^\infty \!\!\!
dk \; k^2 e^{-k^2\sigma^2/2} \, \frac{\sin^2
\frac{E_\bk}{2\hbar}t }{E_\bk\left( \epsilon_\bk + 2 g_B n_0 \right)} \nonumber \\
&\times& \left(2 - 2\frac{\sin 2kL}{2kL} - 2\frac{\sin2kD}{2kD}+\frac{\sin 2k(L+D)}{2k(L+D)}+\frac{\sin
2k(D-L)}{2k(D-L)} \right)  \nonumber\\
& \equiv&  2\Gamma_0(t) + \delta^c(t) \label{gammami}
\end{eqnarray}
In the above equations it is easy to identify the term $\delta^c(t)$ which quantifies 
the deviation to the dechoherence exponent $2\Gamma_0$
typical of the decoherence of two impurities interacting with independent environments. Note that while $\Gamma_0$ depends only on $\bf L$ i.e. 
on the spatial size of the double well, $\delta$ depends non trivially on ${\bf L} \pm {\bf D}$ i.e. on the distance between the impurities of different wells.
As before  the special case of a bath of noninteracting bosons 
$ \Gamma_1^{n.i.}(t)${\bf ,}  $ \Gamma_2^{n.i.}(t)$ are obtained from the above equations
(\ref{gammapl})  simply imposing $g_B=0$ and $E_\bk=\epsilon_\bk$.

\begin{figure}
  \centering
  \includegraphics[height=7cm]{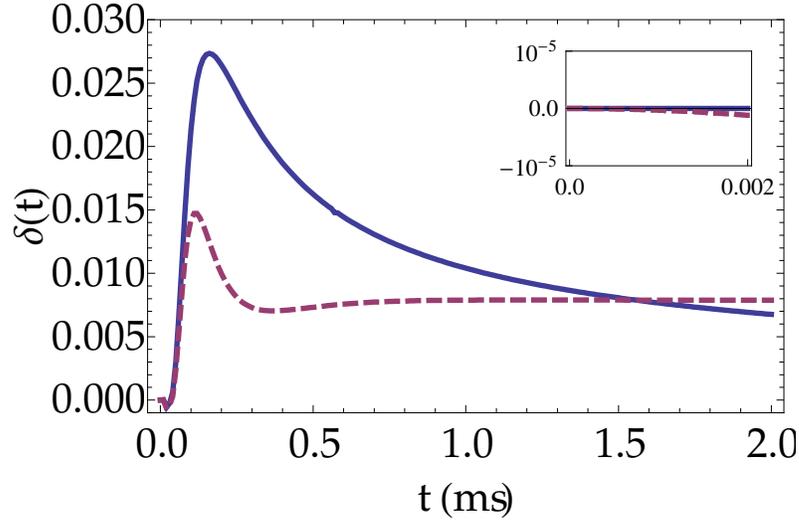}
  \caption{$\delta^c(t)$ (dashed line) and $\delta^{n.i.}(t)$ 
  (solid line) vs. time for a pair of impurity atoms in a three-dimensional environment.
  The inset shows $\delta(t)$ for very short times $0 \le t \le 2 \, \mu \rm{s}$. }
  \label{deltas}
\end{figure}

\begin{figure}
  \centering
  \includegraphics[height=8cm]{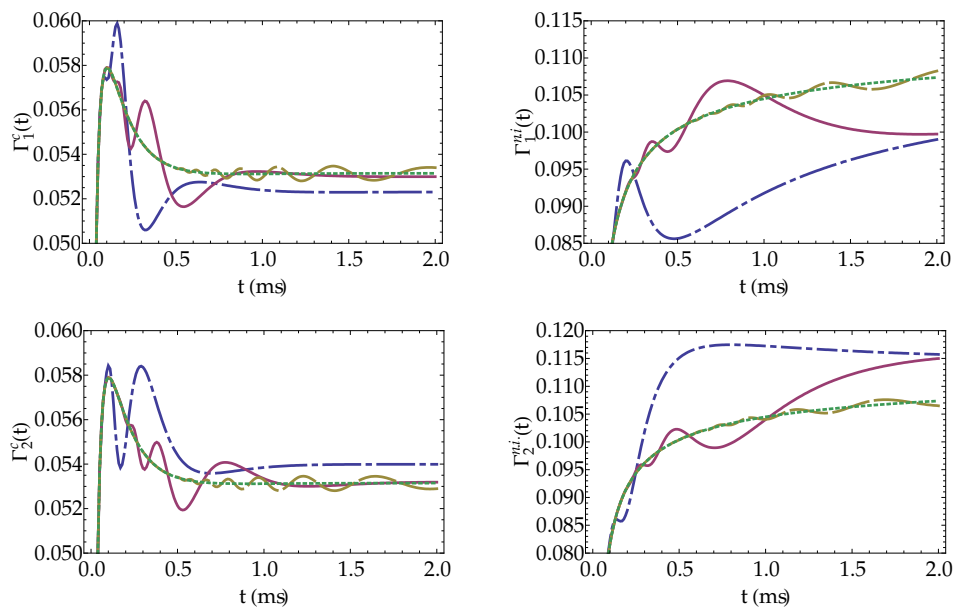}
  \caption{$\Gamma_1(t)$ (top) and $\Gamma_2(t)$ (bottom) vs. time for a pair of impurity atoms interacting with a bosonic condensate (left) and with free bosons (right) in three
  dimensions for different
  distances between the impurities: $2D=8L$ (dash-dotted line),
  $2D=16L$ (solid line), and $2D=40L$
  (dashed line); $2\Gamma_0(t)$ (dotted line) is also shown for comparison.}
  \label{figgammasdistance}
\end{figure}

As in the case of single impurity decoherence the impurities do not loose all their coherence: $\Gamma_1$ and $\Gamma_2$
saturate to a stationary value that can be varied with the help of
Feshbach resonances. Furthermore 
Fig.\ref{figgammaboth} shows that in a system of two impurities  coherence oscillations
appear, both for interacting and non--interacting
bosons in the environment (even more pronounced oscillation are shown in Fig.   \ref{figgammasdistance}). Such coherence revival is due to the collective nature of the coupling, as quantified by $\delta^c(t)$ ($\delta^{n.i.}(t)$ for free bosons).
As shown in Fig.\ref{deltas} also the $\delta(t)$'s are characterized by three different time scales comparable to those analysed
for $\Gamma_0(t)$. In the first stage the difference $|\delta(t)|$ is negligible,
since the presence of each impurity cannot have modified yet the environment
seen by the other one; in the second stage, corresponding to the Markovian
dynamics, the difference $|\delta(t)|$ steadily grows up; and in the third stage
it decreases, reaching a stationary value.

For a pair of impurities we observe super -- and sub -- decoherence, however with a peculiarity
which is characteristic of the system here considered. 
Indeed we observe sub -- decoherence in $ \Gamma_1 \equiv \Gamma_{\{0,0\},\{1,1\}} $
and super - decoherence with $ \Gamma_2 \equiv \Gamma_{\{1,0\},\{0,1\}}$,
at variance with what one
observes in a standard spin boson model, where their role would be exchanged \cite{massimo}.
This different behaviour is due to the particular 
configuration of our system: $\Gamma_1$ gets
contribution from superpositions of the states $\ket{0,0}$ and $\ket{1,1}$,
where the atoms sit in wells with identical distance, whereas the states
$\ket{0,1}$ and $\ket{1,0}$, contributing to $\Gamma_2$, correspond to atoms
sitting in wells with different separations.

Further insight on the features of the collective decoherence is gained by considering the 
decoherence of impurities sitting in sites which are at a larger distance than
$2D=4L=600 \, \rm{nm}$. 
In Fig.\ref{figgammasdistance} we plot the decoherence exponents for impurities trapped in lattice sites at distances $2D=8L, \, 2D=16L$,
and $2D=40L$ respectively. 
These plots suggest the following picture: initially the impurities decohere
independently, as if they were each immersed in its own environment; at some later time, the
environment correlations due to the impurities act back on them and give
rise to oscillating deviations from $2\Gamma_0(t)$.
The onset time of these oscillations
depends on the separation: the larger the separation, the later the onset. On the other
hand, the correlations become weaker as the distance increases and
the oscillations become consequently smaller in amplitude.
At large separation (here, approximately $40L$), the parameters $\Gamma_1$ and $\Gamma_2$
are hardly discernible from $2\Gamma_0$, since the environment correlations induced by the
impurities vanish. Similar features in a related context are reported in \cite{hanggi}. 
In summary $\Gamma_1(t)$ and $\Gamma_2(t)$ also prompt information
about the correlation length of the environment.

\subsection{Decoherence in one dimension}

\begin{figure}
  \centering
  \includegraphics[width=16cm]{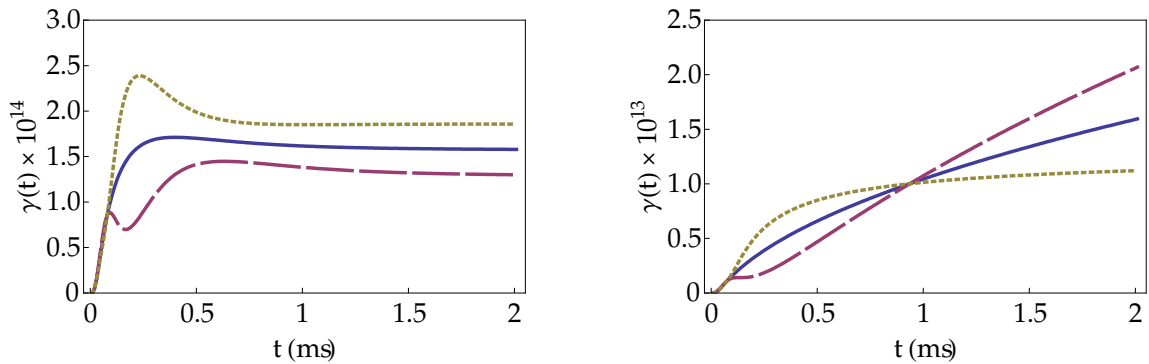}
  \caption{$\gamma_1(t)$ (dashed line), $\gamma_2(t)$ (dotted line), and $2\gamma_0(t)$
  (solid line) vs. time for a pair of impurity atoms immersed in a condensate (left)
  and in an environment of free bosons (right) in one dimension. The separation between
  two impurity atoms is $2D=4L$.}
  \label{figgammaboth1d}
\end{figure}

Finally, we examine the decoherence process in a one-dimensional condensate. Since, as previously discussed, the spectral density (\ref{specden}) is superohmic for an interacting gas, but subohmic for a free Bose gas, we expect qualitative different results for the two cases, in contrast to the three-dimensional case.  
The decay exponents in one dimension $\gamma(t)$  become

\begin{equation}
\gamma_0^c(t) = \frac{4g_{AB}^2 n_0}{\pi}\int_{-\infty}^{\infty} \!\!\!
dk \left[\; e^{-k^2\sigma^2/2}\frac{\sin^2
\frac{E_\bk}{2\hbar}t}{E_\bk\left( \epsilon_\bk + 2 g_B n_0 \right)}
 \right]  \sin^2 kL
 \label{gammaz1d}
\end{equation}
for one impurity and

\begin{eqnarray}
\gamma_1^c(t) & = &  \frac{4g_{AB}^2 n_0}{\pi}\int_{-\infty}^{\infty} \!\!\!
dk \left[\; e^{-k^2\sigma^2/2}\frac{\sin^2
\frac{E_\bk}{2\hbar}t}{E_\bk\left( \epsilon_\bk + 2 g_B n_0 \right)}
 \right]  \sin^2 \left( kL \right) \cos^2 \left( kD \right)  \nonumber\\
&\equiv&  2\gamma_0(t) - \delta^c(t) \label{gammapl1d}\\
\gamma_2^c(t) & = & \frac{4g_{AB}^2 n_0}{\pi}\int_{-\infty}^{\infty} \!\!\!
dk \left[\; e^{-k^2\sigma^2/2}\frac{\sin^2
\frac{E_\bk}{2\hbar}t}{E_\bk\left( \epsilon_\bk + 2 g_B n_0 \right)}
 \right]  \sin^2 \left( kL \right) \sin^2 \left( kD \right) \nonumber\\
& \equiv&  2\gamma_0(t) + \delta^c(t) \label{gammami1d}
\end{eqnarray}
for two impurities in a condensate. The behaviour of these parameters critically
depends on the nature of the environment, see Fig.\ref{figgammaboth1d}.
In particular, decoherence in a one-dimensional sample of free bosons
results Markovian, in agreement with the naive expectation, due to its subohmic spectral density.

\section{Conclusions}
We have shown how a system of impurity atoms trapped in an array of double wells,
interacting with a cold atomic gas, is described, in a suitable regime, 
by a spin - boson hamiltonian. The specific nature of our system, in which the pseudospins, associated 
with the presence of an impurity in the right/left well of each site, have a spatial  dimension introduces peculiar 
features  in the decoherence of a single impurity as well as in the collective decoherence, 
with the persistence of coherence at long times, the presence of coherence oscillations and counterintuitive super / sub decoherent states.

We have shown in particular that for a three-dimensional bath one never has a Markovian behaviour. A one-dimensional bath is in this respect more interesting since one can go from a non-Markovian to a Markovian behaviour just by tuning the interaction of the bath.

As a final comment we would like to say a few words about the role of the quadratic terms in the Bogoliubov
operators which we have neglected in our derivation of Hamiltonian (\ref{hint}). Although a detailed study of their 
effects is beyond the scope of the present article, we would like to point out that their effects are negligible with respect to the
linear terms we have analyzed in the present manuscript.  One can show that their inclusion amounts to taking into account 
elastic scattering of Bogoliubov particles, which is simply responsible of an energy shift, inelastic scattering processes  and Bogoliubov pair creation and annihilation. In these two latter additional terms the length of wave vectors  $k$  that can play some role in the impurities' dynamics 
is limited from below by the finite  size of the condensate and from above by cutoff parameter $\sigma^{-1}$. It can be shown that, in this frequency range,  the coupling constants of the neglected processes are, for the values of parameters assumed in our analysis, three orders of magnitude smaller than the coupling constants $\hbar \Omega_{n,{\bf k}}^i $ of the linear terms. As a consequence, a rough estimate leads us to suppose that any possible relevant effect 
of the quadratic terms in the Hamiltonian would become apparent at time scales that are
three orders of magnitude larger than those examined in this article.

\ack
The authors acknowledge financial support from MIUR through
the project PRIN 2006 "Collective decoherence in engineered reservoirs" and from EUROTECH S.p.A.;
GDC is supported by the Spanish Ministry of Science and Innovation through the program Juan de la Cierva. AR acknoledges support also by the EuroQUAM
FerMix program.

\appendix
\section{Disentangling the time evolution operator}
The factorization of the time evolution operator
$\hat{U}(t) = \exp \left[ -i\hat{H}t/\hbar \right]$ is often an impossible
task. When the Hamiltonian contains operators forming a
Lie algebra the transformation of $\hat{U}(t)$ into a product
of simpler exponential operators is however possible in some cases \cite{disen}.
Here we show a practical way to transform $\hat{U}(t)$, which we write as

\begin{eqnarray}
\hat{U}(t) & = & \exp \left[ -\frac{i}{\hbar}\sum_\bk E_\bk \hcdag_\bk
\hc_\bk t \right] \exp \left[ \sum_\bk \left( \sum_i A_\bk^i(t)\sigz^i+
\alpha_\bk^{}(t) \right) \hcdag_\bk
 \right]
\nonumber \\
& & \times \exp \left[ - \sum_\bk \left( \sum_i B_\bk^i(t)\sigz^i+
\beta_\bk^{}(t) \right) \hc_\bk
 \right]\hat{U}_{R}(t)
 \label{ufact}
\end{eqnarray}
where $\hat{U}_{R}(t)$ is to be determined, as well as the quantities
$A_\bk^i(t),B_\bk^i(t),\alpha_\bk(t)$ and $\beta_\bk(t)$. Since at $t=0$
the time evolution operator $\hat{U}$ reduces to the identity operator,
$A_\bk^i(0)=B_\bk^i(0)=\beta_\bk(0)=\alpha_\bk(0)=0$. All unknown quantities can be found
with the help of the relation

\begin{equation}
\hat{H} = i\hbar \left[d\hat{U}(t)/dt \right] \hat{U}^{-1}(t)
\label{match}
\end{equation}
which holds for any time-independent Hamiltonian and of the relation

\begin{equation}
e^{\hat{X}} \hat{Y} e^{-\hat{X}}=
\hat{Y}+\left[ \hat{X},\hat{Y} \right]+
\frac{1}{2}\left[ \hat{X},\left[ \hat{X}, \hat{Y} \right] \right]
+\frac{1}{6}\left[ \hat{X},\left[ \hat{X}, \left[ \hat{X}, \hat{Y} \right] \right] \right] + \ldots
\end{equation}
for arbitrary operators $\hat{X}$ and $\hat{Y}$.
After inserting the expression (\ref{ufact}) for the time evolution operator $\hat{U}(t)$
in the right-hand-side of (\ref{match}), a comparison with the Hamiltonian
(\ref{hpseudo}) leads to the expressions

\begin{equation}
A_\bk^i(t) = \frac{\hbar
\left(\Omega_{R,\bk}^{i*}-\Omega_{L,\bk}^{i*}
\right)}{2E_\bk}\left( 1-e^{iE_\bk t/\hbar} \right), \;\;\;
B_\bk^i(t)=A_\bk^{i*}(t),
\end{equation}

\begin{equation}
\alpha_\bk^{}(t)= \frac{\hbar \sum_i
\left(\Omega_{R,\bk}^{i*}+\Omega_{L,\bk}^{i*}
\right)}{2E_\bk}\left( 1- e^{iE_\bk t/\hbar}\right), \;\;\;\;\;
\beta_\bk^{}(t)=\alpha_\bk^{*}(t)
\end{equation}
for $A(t)$, $B(t)$, $\alpha(t)$ and $\beta(t)$,
and to the differential equation

\begin{eqnarray}
\frac{d}{dt}\hat{U}_R(t) & = & -\sum_\bk \left( \sum_i \dot{B}_\bk^i(t) \sigz^i
+\dot{\beta}_\bk^{}(t) \right) \left( \sum_j
A_{\bk}^j(t) \sigz^j +\alpha_{\bk}^{}(t)\right)\hat{U}_R(t)
\label{diff}
\end{eqnarray}
for the unknown exponential operator $\hat{U}_R(t)$, which we write as

\begin{eqnarray}
\hat{U}_R(t) & = &  \exp \left[ -\sum_{\bk} \left(
\sum_{ij}\eta_{\bk}^{ij}(t)\sigz^i\sigz^j+\sum_{i}\mu_{\bk}^{i}(t)\sigz^i+\epsilon_{\bk}(t)\right) \right]
\end{eqnarray}
A comparison with (\ref{diff}) gives

\begin{equation}
\dot{\eta}_{\bk}^{ij}(t)= \dot{B}_\bk^i(t)A_{\bk}^j(t),
\;\;\; \dot{\epsilon}_{\bk}(t)= \dot{\beta}_\bk(t)\alpha_{\bk}(t),
\;\;\; \dot{\mu}_{\bk}^{i}(t)= \dot{B}_\bk^i(t)\alpha_{\bk}(t)+
\dot{\beta}_\bk^{}(t)A_{\bk}^i(t)
\end{equation}
i.e.,

\begin{eqnarray}
\eta_{\bk}^{ij}(t) & = &
-i\hbar\frac{\left(\Omega_{R,\bk}^{i}-\Omega_{L,\bk}^{i}\right)
\left(\Omega_{R,\bk}^{j*}-\Omega_{L,\bk}^{j*}\right)}{4E_{\bk}}
\left[
t+\frac{i\hbar}{E_\bk}\left(1-e^{-iE_\bk t/\hbar}\right)\right] \\
\epsilon_{\bk}(t) & = &
-i\hbar\frac{\sum_{ij}\left(\Omega_{R,\bk}^{i}+\Omega_{L,\bk}^{i}\right)
\left(\Omega_{R,\bk}^{j*}+\Omega_{L,\bk}^{j*}\right)}{4E_{\bk}}
\left[
t+\frac{i\hbar}{E_\bk}\left(1-e^{-iE_\bk t/\hbar}\right)\right] \\
\mu_{\bk}^{i}(t) & = & -\frac{i\hbar}{2E_\bk}
\Re\left[\left(\Omega_{R,\bk}^{i}-\Omega_{L,\bk}^{i}\right)
\sum_j\left(\Omega_{R,\bk}^{j*}+\Omega_{L,\bk}^{j*}\right)
\right]\left[ t+\frac{i\hbar}{E_\bk}\left(1-e^{-iE_\bk
t/\hbar}\right)\right]
\end{eqnarray}
Moreover, using Glauber's relation

\begin{eqnarray}
\exp \left[ \sum_\bk g_\bk^{} \hcdag_\bk \right] \exp \left[
-\sum_\bk g_\bk^* \hc_\bk \right] = \exp \left[ \sum_\bk \left(
g_\bk^{} \hcdag_\bk - g_\bk^* \hc_\bk \right) \right] \exp
\left[\frac{1}{2}\sum_\bk \left| g_\bk \right|^2 \right]
\label{glau}
\end{eqnarray}
the two exponentials linear in Bogoliubov operators can be merged into

\begin{eqnarray}
& & \exp \left[ \sum_\bk \left( \sum_i A_\bk^i(t) \sigz^i+
\alpha_\bk (t) \right)\hcdag_\bk  \right]
\exp \left[ -\sum_\bk \left( \sum_i B_\bk^i(t) \sigz^i+\beta_\bk(t) \right) \hc_\bk \right] \\
& &= \exp \left\{ \left[ \sum_\bk \left( \sum_i A_\bk^i(t)
\sigz^i+ \alpha_\bk (t) \right)\hcdag_\bk - \sum_\bk \left( \sum_i
A_\bk^{i*}(t) \sigz^i+ \alpha_\bk^* (t) \right)\hc_\bk \right]
\right\} \nonumber \\
& & \times \exp \left\{\frac{1}{2} \left[ \sum_\bk \left( \sum_i
A_\bk^i(t) \sigz^i+ \alpha_\bk (t) \right) \left( \sum_j
A_\bk^{j*}(t) \sigz^j+ \alpha_\bk^* (t) \right) \right] \right\}
\end{eqnarray}
and the contribution of the last exponential can be included in $U_R(t)$.
Performing some commutations where it is possible,
the time evolution operator becomes

\begin{eqnarray}
\hat{U}(t) & = & \exp \left[ -\frac{i}{\hbar}\sum_\bk E_\bk \hcdag_\bk
\hc_\bk t \right] \exp \left[ -\sum_{\bk} \left(
\sum_{ij}\eta_{\bk}^{ij}(t)\sigz^i\sigz^j+\sum_{i}\mu_{\bk}^{i}(t)\sigz^i+\epsilon_{\bk}(t)\right) \right]
 \nonumber \\
 & & \times \exp \left[ \sum_\bk \left( \sum_i A_\bk^i(t)\sigz^i+
\alpha_\bk^{}(t) \right) \hcdag_\bk  - \sum_\bk \left( \sum_i B_\bk^i(t)\sigz^i+
\beta_\bk^{}(t) \right) \hc_\bk \right] \nonumber \\
& & \times \exp \left\{\frac{1}{2} \left[ \sum_\bk \left( \sum_i
A_\bk^i(t) \sigz^i+ \alpha_\bk (t) \right) \left( \sum_j
A_\bk^{j*}(t) \sigz^j+ \alpha_\bk^* (t) \right) \right] \right\}
\end{eqnarray}
Finally, the exponential operators that do not contain bath
operators commute, so the time evolution operator can be further
modified into the final form (\ref{equt}).

\section{Derivation of the dynamics of the impurities}

The action of $\hat{U}_{\bk,D}(t)$ on a pure state of the
whole system is

\begin{eqnarray}
& & \hat{U}_{\bk,D}(t) \ket{\{n_p\}}\bra{\{m_p\}} \otimes \rho_{\bk}^{} \hat{U}_{\bk,D}^{\dagger}(t)=
\ket{\{n_p\}}\bra{\{m_p\}} \otimes \nonumber \\
& & \exp \left[ \left( -\sum_j A_\bk^j(t)(-1)^{n_j}+\alpha_\bk(t)
\right)\hcdag_\bk -
 \left(-\sum_j A_\bk^{j*}(t)(-1)^{n_j}+\alpha_\bk^*(t) \right) \hc_\bk
\right] \rho_{\bk}^{} \nonumber \\
& & \exp \left[ -\left( -\sum_j A_\bk^j(t)(-1)^{m_j}+\alpha_\bk(t)
\right)\hcdag_\bk +
 \left(-\sum_j A_\bk^{j*}(t)(-1)^{m_j}+\alpha_\bk^*(t) \right) \hc_\bk
 \right]
\end{eqnarray}
and the density matrix elements $\rho_{\{n_p\},\{m_p\}}(t)$ of the impurities are
obtained by tracing over the bath,

\begin{eqnarray}
\rho_{\{n_p\},\{m_p\}}(t) & = & \exp \left\{i \Theta_{\{n_p\},\{m_p\}}(t) \right\}
\exp \left\{i \Xi_{\{n_p\},\{m_p\}}(t) \right\}\rho_{\{n_p\},\{m_p\}}(0) \nonumber \\
& & \times \bra{\{n_p\}} \prod_\bk \mbox{Tr}_{B,\bk} \left\{ \hat{U}_{\bk,D}(t)
\ket{\{n_p\}}\bra{\{m_p\}} \otimes \rho_{\bk}^{}
\hat{U}_{\bk,D}^{\dagger}(t) \right\} \ket{\{m_p\}}
\end{eqnarray}
where $\mbox{Tr}_{B,\bk}$ denotes the trace over each Bogoliubov mode of the
environment and the phases

\begin{eqnarray}
\Theta_{\{n_p\},\{m_p\}}(t) & = & \hbar^2 \sum_\bk \frac{f_\bk(t)}{4E_\bk^2}
\nonumber \\
& & \times \sum_{ij}\Re \left(\Omega_{R,\bk}^{i}-\Omega_{L,\bk}^{i}\right)
\left(\Omega_{R,\bk}^{j*}-\Omega_{L,\bk}^{j*}\right)\left[ (-1)^{n_i+n_j}-(-1)^{m_i+m_j} \right] \\
\Xi_{\{n_p\},\{m_p\}}(t) & = & \hbar^2 \sum_\bk \frac{f_\bk(t)}{E_\bk^2}\Re
\sum_j\left(\Omega_{R,\bk}^{j*}+\Omega_{L,\bk}^{j*}\right)
\sum_i\left(\Omega_{R,\bk}^{i}-\Omega_{L,\bk}^{i}\right)
\left( n_i-m_i \right)
\end{eqnarray}
come from the unitary operators in (\ref{equt}).
Performing ciclic permutation inside the trace and using the
identity $\exp (\hat{M}) \exp (\hat{N}) = \exp (\hat{M}+\hat{N})
\exp [\hat{M},\hat{N}]/2$, which holds for
operators $\hat{M}$ and $\hat{N}$ that commute with their commutator, the trace
$\mbox{Tr}_{B,\bk}$ in (\ref{eqrho}) becomes

\begin{eqnarray}
& & \exp \left[ i \Im \left( \sum_j
A_\bk^j(t)(-1)^{n_j}-\alpha_\bk(t) \right)
 \left(\sum_j A_\bk^{j*}(t)(-1)^{m_j}-\alpha_\bk^*(t) \right)\right\}
\nonumber \\
& & \times \mbox{Tr}_{B,\bk}\left\{ \exp \left[ 2\sum_i
\left(n_i-m_i\right) \left( A_\bk^i(t)\hcdag_\bk - A_\bk^{i*}(t)
\hc_\bk \right) \right] \rho_{\bk}^{} \right\} \nonumber \\
& & \equiv \exp \left\{i \Delta_{\{n_p\},\{m_p\}}(t) \right\}
\mbox{Tr}_{B,\bk}\left\{ \exp \left[ 2\sum_i
\left(n_i-m_i\right) \left( A_\bk^i(t)\hcdag_\bk - A_\bk^{i*}(t)
\hc_\bk \right) \right] \rho_{\bk}^{} \right\}
\end{eqnarray}
The trace over the thermal bath of the displacement operators is
well known \cite{massimo},

\begin{eqnarray}
\mbox{Tr}_{B,\bk} \left[ \exp \left\{g_\bk \hcdag_\bk - g_\bk^*
\hc_\bk \right\}\rho_{\bk}\right] & = & \exp \left\{ -
\frac{|g_\bk|^2}{2}\coth \frac{\beta E_\bk}{2} \right\}
\end{eqnarray}
where $\beta=(k_BT)^{-1}$, and leads to equation (\ref{eqrho}).

\section{The coupling constant in a deep optical lattice}

In a deep optical lattice the ground state wave
functions of each well can be approximated with those of harmonic oscillators,

\begin{eqnarray}
\varphi_{i,N}(\bx) & = & \frac{1}{\left[\pi^3 x_0^2 y_0^2
z_0^2\right]^{1/4}} \exp \left[ -\frac{(x-x_{i,N})^2}{2
x_0^2}-\frac{(y-y_{i,N})^2}{2 y_0^2} -\frac{(z-z_{i,N})^2}{2
z_0^2} \right].
\end{eqnarray}
Here $N=L,R$, and $x_0=\sqrt{\hbar/(m\omega_x)}$,
$y_0=\sqrt{\hbar/(m\omega_y)}$, and
$z_0=\sqrt{\hbar/(m\omega_z)}$, where the $\omega$'s are the
trapping frequencies of the harmonic trap approximating the
lattice potential at bottom of L and R wells of the lattice site
$i$. The coupling frequencies (\ref{ome}) of the spin-boson model then become

\begin{eqnarray}
\Omega_{n,\bk}^i & = & \frac{g_{AB}\sqrt{n_0}}{\hbar}\left(
|u_\bk|-|v_\bk| \right) \int d^3x \; |\varphi_{i,L}(\bx)|^2 e^{i
\bk \cdot \bx} \nonumber \\
& = & \frac{g_{AB}\sqrt{n_0}}{\hbar}\left(
|u_\bk|-|v_\bk| \right) e^{-k^2\sigma^2/4}e^{ik_x x_{i,n}}, \;\;\;\;\; n=L,R
\end{eqnarray}
having assumed identical confinement in the three directions,
$\sigma=x_0=y_0=z_0$.

\end{document}